# Convolutional Neural Networks Towards Facial Skin Lesions Detection

Reza Sarshar
*School of Industrial and Systems Engineering*
*Tarbiat Modares University*
Tehran, Iran
reza_sarshar@modares.ac.ir

Mohammad Heydari
*School of Industrial and Systems Engineering*
*Tarbiat Modares University*
Tehran, Iran
m_heydari@modares.ac.ir

Elham Akhondzadeh Noughabi
*School of Industrial and Systems Engineering*
*Tarbiat Modares University*
Tehran, Iran
elham.akhondzadeh@modares.ac.ir

*Abstract*— Facial analysis has emerged as a prominent area of research with diverse applications, including cosmetic surgery programs, the beauty industry, photography, and entertainment. Manipulating patient images often necessitates professional image processing software. This study contributes by providing a model that facilitates the detection of blemishes and skin lesions on facial images through a convolutional neural network and machine learning approach. The proposed method offers advantages such as simple architecture, speed and suitability for image processing while avoiding the complexities associated with traditional methods. The model comprises four main steps: area selection, scanning the chosen region, lesion diagnosis, and marking the identified lesion. Raw data for this research were collected from a reputable clinic in Tehran specializing in skincare and beauty services. The dataset includes administrative information, clinical data, and facial and profile images. A total of 2300 patient images were extracted from this raw data. A software tool was developed to crop and label lesions, with input from two treatment experts. In the lesion preparation phase, the selected area was standardized to 50×50 pixels. Subsequently, a convolutional neural network model was employed for lesion labeling. The classification model demonstrated high accuracy, with a measure of 0.98 for healthy skin and 0.97 for lesioned skin specificity. Internal validation involved performance indicators and cross-validation, while external validation compared the model's performance indicators with those of the transfer learning method using the Vgg16 deep network model. Compared to existing studies, the results of this research showcase the efficacy and desirability of the proposed model and methodology.

*Keywords— Deep Learning, Convolutional Neural Networks, Skin Lesions, Image Processing*

## I. Introduction

A person's face is pivotal in shaping their social interactions, providing a wealth of information including identity, gender, race, age, emotions, health, personality, and attractiveness. Recent advancements in computer and digital camera performance have propelled the field of computer-based face image analysis into the spotlight. Facial recognition, the most extensively researched aspect within this field, has proven successful in applications such as access control, internal security, and computer security [1].

However, the research of computer-based face analysis is still in its early stages, with numerous challenges awaiting discovery [2]. Visual examination of the facial skin surface for identifying lesions that raise clinical suspicions of skin cancer or pigmented abnormalities is a time-consuming process known as external skin lesion screening [3].
The characteristic assessment of skin lesions, encompassing features like asymmetry, border roughness, color distribution, diameter, and deformity, forms the foundation of initial screening. These visual descriptors play a crucial role in aiding dermatologists to identify and evaluate skin lesions [4]. Recent studies highlight the association between access to skin screenings and early diagnosis, leading to improved prognoses [5][6][7].

As an illustrative example, acne, ranking as the eighth most prevalent skin disorder globally, is typically assessed for severity by dermatologists in a clinical setting. A skin lesion, defined as a change in surface color or texture, such as an erosion or ulcer, can manifest in various parts of the body [8]. However, its appearance on the face can have adverse effects on a person's self-image. For instance, acne, often considered a disorder among young individuals, is increasingly affecting those over twenty-five years old, presenting both a medical concern and raising cosmetic apprehensions [9].

In contemporary digital cameras, there are skin enhancement functions specifically designed for capturing human images. These functions excel at adjusting the brightness and color of human skin, often surpassing the natural state. Many individuals prefer utilizing these functions when taking pictures. In the medical field, leveraging images for the diagnosis of conditions like acne and skin spots is imperative for effective treatment [10].
Despite recent technological advancements in smartphones, the availability of high-quality personal cameras, and powerful computing systems, the utilization of images from personal devices for clinical diagnosis remains relatively uncommon. This is attributed in part to challenges associated with feature extraction and accurate classification in the presence of image artifacts.

This research proposes a skin feature learning framework that integrates both lesion and healthy skin contexts, facilitating a comprehensive detection of facial skin lesions.

The method's advantage lies in its ability to overcome the complexities of image processing, including challenges posed by diverse skin lesions, variations in facial angles, camera distance, light intensity, color intensity, and the presence of artifacts such as cosmetics in images. This reduction in complexity aims to assist doctors and patients in visual screening for the presence of lesions on the face.

The paper structure unfolds as follows: in the second part, the background of the subject and methods of skin lesion diagnosis are presented. The third part details the research methodology and materials, while the fourth part delves into the development of the proposed research ranking model. Moving on to the fifth section, we scrutinize the numerical results and validation of the model. The sixth section addresses the practical implementation of the model in the real world. Finally, in the seventh section, conclusions are shown, and proposals for further study are offered.

## II. RELATED WORKS

The classification of skin lesions stands as a crucial stage in the computerized analysis of dermoscopic images. Misclassification can have adverse effects on subsequent steps in an automated computer-aided skin lesion detection system. Given the gradual spread of skin lesions to adjacent tissues, there exists a complex relationship within various sections of the anatomical structure of these lesions. Leveraging the correlation between candidate pixels and their surrounding areas for machine learning serves as a valuable distinguishing feature.

Historically, various methods have been proposed for lesion diagnosis. Some of these approaches aimed to identify lesions through different image processing methods, including threshold definition, texture analysis, brightness intensity, desired area, and color segmentation, either individually or in combination with non-deep machine learning approaches [11][12][13][14][15][16][17]. However, these methods faced complications and limitations related to image processing in machine learning. Consequently, recent research has shifted its focus towards utilizing machine learning methods based on deep neural networks.

For instance, Oliveria [18] employed a conditional adversarial neural network classifier to train images of healthy and lesional skin, achieving an accuracy of 88.33% in distinguishing between the two.

In 2019, Wang et al. [19] introduced a two-way skin feature learning framework, implemented atop a deep convolutional network, ultimately detecting skin lesions with an accuracy of 85.35%.

In 2021, Soensken et al. [20] offered a deep neural network model identifying suspicious cancerous lesions in images captured by personal cameras on the skin surface. This model, based on a dataset of 133 patients, classified images related to skin lesions into 6 classes, achieving a sensitivity of 90.3% and specificity of 89.9%. While results from these studies and others on deep learning indicate that models based on deep learning can achieve comparable or superior diagnostic accuracy in specific visual tasks compared to dermatologists, challenges persist in real-world applications. For instance, in the mentioned research, there is a notable absence of attention to the efficiency of the offered method in detecting the presence of a lesion on the face without the need for detailed classification of the lesion type, adding to the complexity of model implementation.

## III. RESEARCH METHODOLOGY

The research comprises four distinct stages:

*1) Database Creation and Patients Information Extraction*

In this initial stage, patient data from a reputable medical center specializing in skin and beauty services were collected. Following several refinement stages, 2300 patient files were selected for further research.

*2) Software Design for Image Cutting and Tagging*

A dedicated software with a user interface was designed and implemented for image cutting and labeling, as illustrated in Fig. 1. In this phase, a medical expert utilized the software to precisely cut the lesion area and healthy skin on the face, producing images of dimensions 50x50 pixels. The labeled images were reviewed and verified by another expert.

*3) Image Preprocessing and CNN Model Design*

The acquired image data underwent preprocessing and normalization using pre-existing OpenCV libraries. Subsequently, a multi-layer model named DisorderNet, based on a convolutional neural network, was developed using TensorFlow and Keras libraries. The training process involved 70% of the data, while 20% was allocated for validation, and the remaining 10% for model testing. Internal validation utilized performance indicators and cross-validation, while external validation involved comparing the model's performance indicators with those of the transfer learning method, leveraging the VGG16 deep network model pretrained with the ImageNet dataset [21].

*4) Sliding Window Implementation for Lesion Detection*

In the final stage, a sliding window mechanism was implemented to traverse the face, facilitating lesion detection, and marking. This step enhances the model's capability to detect lesions across various facial regions. This systematic approach ensures a comprehensive investigation, starting from data collection and refinement, through software design and image processing, to the development and validation of a convolutional neural network model for effective skin lesion detection. The inclusion of external validation with transfer learning methods further enhances the robustness and generalizability of the proposed model.

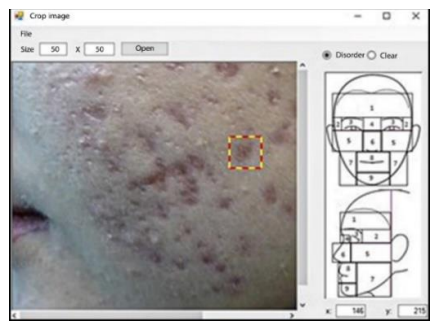

Fig. 1. User interface Developed for cropping and tagging images.

For simplicity, the data has been classified into two classes: "clean skin" and "damaged skin." An illustrative example of the image cuts made can be observed in Fig. 1.

*A. Convolution*

Convolution is a straightforward process involving the application of a matrix, also known as a kernel or filter, to an image. This process can either shrink the image or include additional border layers to maintain its original size. The main purpose of convolution is to extract specific features from the image, such as shapes, edges, etc. This technique finds widespread application in image processing, particularly in CNN and face recognition [22].

*B. Model Contruction*

The model architecture in this research, referred to as DisorderNet, is composed of 4 deep convolution layers, each followed by a MaxPooling layer with the Rectified Linear Unit (ReLU) activation function. To mitigate overfitting, the drop trick was employed, randomly removing certain grid cells. The input to the model consists of 50x50 images with 3 color channels (red, green, and blue) in the RGB color space, and the model was trained over 100 epochs. Convolutional Neural Networks (CNNs) are adept at extracting image features, and a deeper CNN is generally capable of capturing more intricate and detailed features.

TABLE I. ARCHITECTURE OF CNN LAYERS

| Layer Type | Output Shape | Parameters |
|---|---|---|
| Conv2D | (48, 48, 32) | 896 |
| MaxPooling2D | (24, 24, 32) | 0 |
| Conv2D | (22, 22, 64) | 18496 |
| MaxPooling2D | (11, 11, 64) | 0 |
| Conv2D | (9, 9, 128) | 73856 |
| MaxPooling2D | (4, 4, 128) | 0 |
| Conv2D | (2, 2, 128) | 147584 |
| MaxPooling2D | (1, 1, 128) | 0 |
| Flatten | (128) | 0 |
| Dropout | (128) | 0 |
| Dense | (512) | 66048 |
| Dense | (1) | 513 |

The number of epochs was considered 100. Fig. 2. depicting changes in loss is presented. The reduction of training cost in higher epochs was shown, and Fig. 3. also shows that as we get closer to higher epochs, the accuracy of the model increases. Additionally, illustrates the variations in the accuracy of the model throughout the learning process. fluctuations in the graph can be caused by outliers. Despite this, the numerical values obtained affirm the validity of the model.

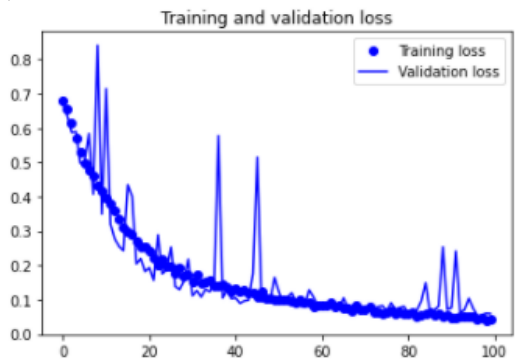

Fig. 2. Loss function during the training process

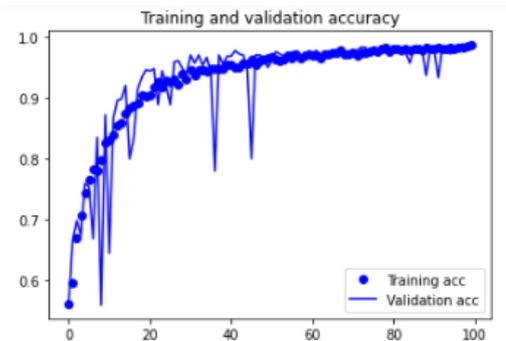

Fig. 3. Accuracy function during the training process

*C. Model Evaluation*

In this section, various performance indicators were utilized to assess the effectiveness of the research ranking model. These indicators encompass sensitivity, specificity, accuracy, and precision. Sensitivity gauges the proportion of actual positives correctly identified, while specificity measures the proportion of actual negatives correctly identified. Accuracy provides an overall assessment of classifier performance, and precision measures the positive predictive value.

Several terms commonly associated with these indicators include True Positive (TP), True Negative (TN), False Positive (FP), and False Negative (FN). FP occurs when a lesion is incorrectly classified as a lesion, while TN denotes healthy skin correctly classified as healthy. TP and TN signify a positive alignment between model detection and real conditions. On the other hand, FP indicates healthy skin erroneously classified as a lesion, and FN signify lesions wrongly classified as healthy skin. FP and FN point to ranking results that deviate from real conditions.

The measured numerical criteria obtained from implementing the model on the test data are presented in TABLE II, while the results from the validation data are outlined in TABLE III.

TABLE II. TEST DATA

| | | *TP* | *FP* | *TN* | *FN* |
|---|---|---|---|---|---|
| | | 118 | 3 | 139 | 0 |
| Sensitivity | Specificity | Precision | Accuracy | Recall | F1 Score |
| 1 | 0.97 | 0.97 | 0.98 | 0.97 | 0.97 |

TABLE III. VALIDATION DATA

| | | *TP* | *FP* | *TN* | *FN* |
|---|---|---|---|---|---|
| | | 234 | 3 | 298 | 5 |
| Sensitivity | Specificity | Precision | Accuracy | Recall | F1 Score |
| 0.97 | 0.99 | 0.98 | 0.98 | 0.98 | 0.98 |

The numerical results obtained highlight the effectiveness of the model settings and the classification approach in the research. The ROC curve is a valuable visual tool for comparing two ranking models. Originating from signal detection theory developed during World War II for analyzing radar images, the ROC illustrates the relationship between the TPR and the FPR. The AUC is a metric used to evaluate the validity of the model. An AUC value closer to 0.5 indicates lower accuracy, while a value closer to 1 signifies higher accuracy.

In the validation of the research ranking model, three stages of cross-validation were conducted with values of k=5, k=6, and k=7. The results, along with the ROC and AUC diagram for k=5, are presented in TABLE IV and Fig. 4. to (8). These visualizations provide a general understanding of the model's performance and its capability to discriminate between positive and negative instances.

TABLE IV. MODEL CROSS-VALIDATION WITH DIFFERENT K VALUES

| | K = 5 | | | | | AUC |
|---|---|---|---|---|---|---|
| **Round1** | Train Data | | | | Test Data | 0.99 |
| | fold1 | fold2 | fold3 | fold4 | fold5 | |
| **Round2** | Train Data | | | | Test Data | 0.98 |
| | fold5 | fold1 | fold2 | fold3 | fold4 | |
| **Round3** | Train Data | | | | Test Data | 1 |
| | fold4 | fold5 | fold1 | fold2 | fold3 | |
| **Round4** | Train Data | | | | Test Data | 0.96 |
| | fold3 | fold4 | fold5 | fold1 | fold2 | |
| **Round5** | Train Data | | | | Test Data | 0.98 |
| | fold2 | fold3 | fold4 | fold5 | fold1 | |
| **Mean of AUC for K=5** | | | | | | **0.98** |

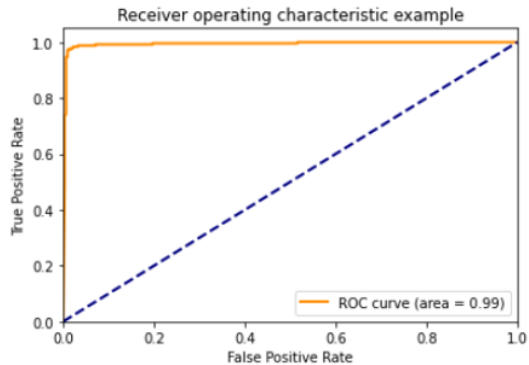
Fig. 4. – ROC in Round1

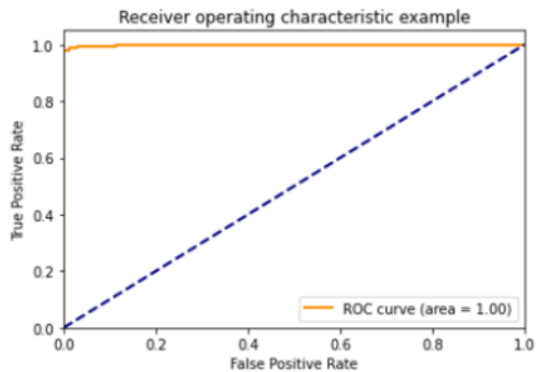
Fig. 5. ROC in Round2

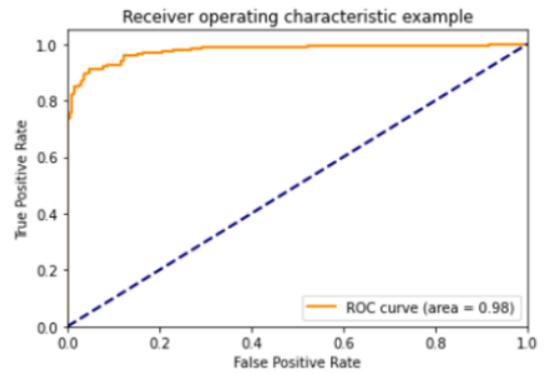
Fig. 6. ROC in Round3

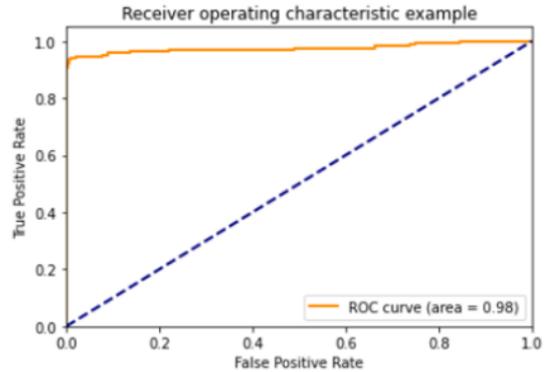
Fig. 7. ROC in Round4

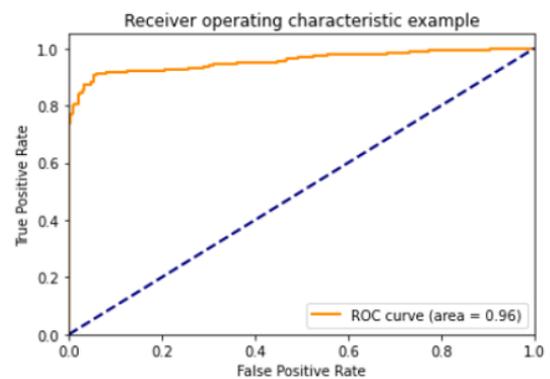
Fig. 8. ROC in Round5

### D. VGG16 TRANSFER LEARNING MODEL

In applying the transfer learning approach, the weights of the initial layers of the Vgg16 network, which were pretrained, were retained without modification. Subsequently, a flat layer was added to the network, and its activation function was set to sigmoid to accommodate the two desired classes. The network was then retrained. An SGD optimizer was employed with a learning rate of 0.001 and a momentum term of 0.9. No other parameters in the model were manually adjusted. Training was initiated with 100 epochs. The validation metrics are detailed in TABLE V, and the ROC and AUC charts are presented in Fig. 9. These visualizations offer insights into the model's performance when utilizing transfer learning for the specified classification task.

TABLE V. VGG16 VALIDATION DATA

| Sensitivity | SPECIFICITY | ACCURACY | PRECISION |
|---|---|---|---|
| 0.96 | 0.98 | 0.97 | 0.97 |

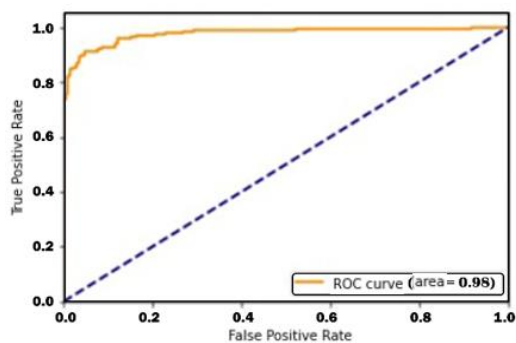

Fig. 9. ROC in Vgg16

The numerical results obtained from the DisorderNet network and its comparison with the transfer learning approach demonstrate the effectiveness and desirability of the research classification. This suggests that the DisorderNet architecture, along with its training methodology, performs well in the classification of skin lesions, and the comparison with transfer learning provides valuable insights into the model's performance and potential strengths.

### E. MODEL EXECUTION IN REAL WORLD

In the real-world implementation of the algorithm, ethical principles were adhered to by selecting a portion of a patient's face. The algorithm, along with the research ranking model, was then applied to this selected facial region. The outcome of this process is depicted in Fig. 10.

In the concluding stage of the research, an algorithm was developed to facilitate the application of the model on a patient's facial image, enabling the identification and marking of skin lesions on the face. The procedural steps for this algorithm are illustrated in Fig. 11.

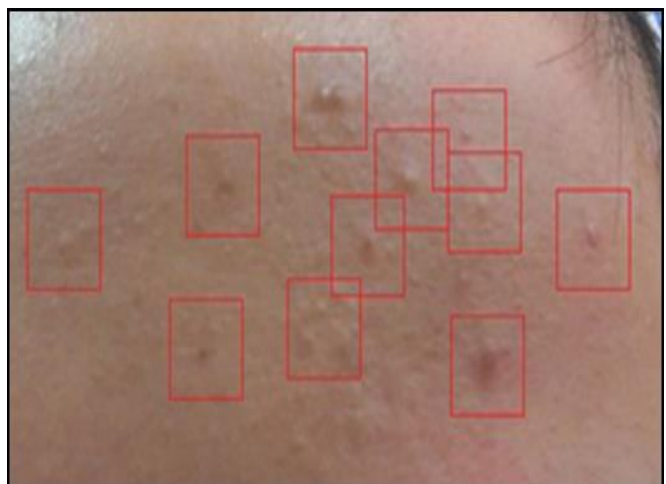

Fig. 10. Diagnosis of a lesion on a part of the patient's face

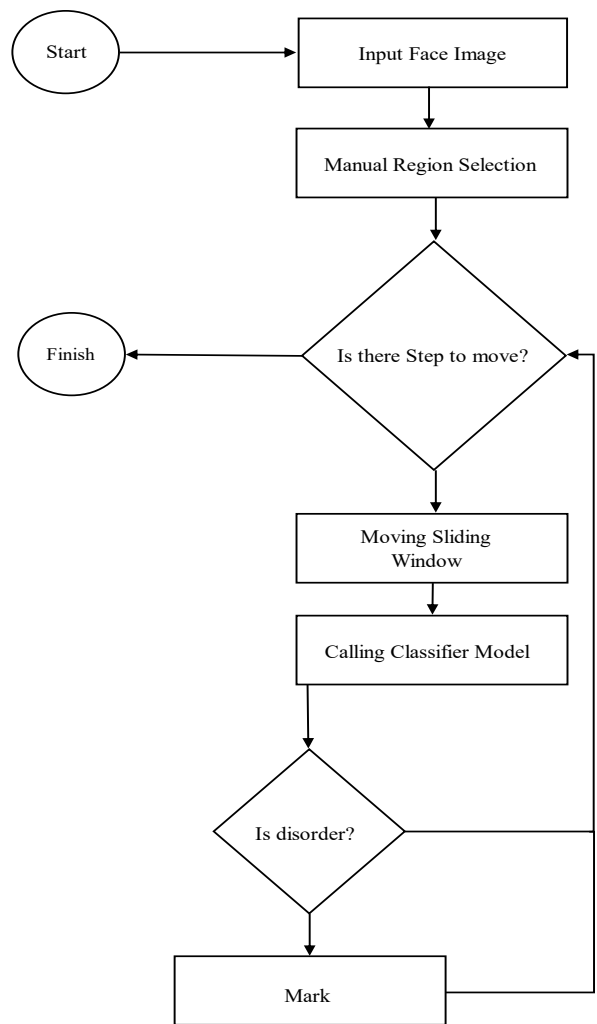

Fig. 11. Proposed Model Flowchart

### F. CONCLUSION

The primary objective of this article was to show the effectiveness of the classification model in detecting skin lesions on the face using a CNN. The model presented here is both straightforward and avoids the complexities often associated with image processing. The high accuracy of 0.98 underscores the robustness and efficacy of the introduced ranking model.

To validate the research, diverse images were utilized, incorporating imaging conditions expected in a medical center. The outcomes of this study hold potential applications in smartphones, medical centers, and entertainment programs. However, it is noteworthy this research did not mention the use of images with varying skin tones, which could be a consideration in future investigations. Additionally, for further automation, future research may explore the inclusion of a non-skin category, alongside the existing categories of healthy and damaged skin.